# A diffraction paradox: An unusually broad diffraction background signals ideal graphene

S. Chen[1,2] M. Horn von Hoegen[4], P. A.
Thiel[1,3] and M.C.Tringides[**,1,2]

[1]*Ames Laboratory - U.S. Department of Energy,* [2]*Department of Physics and Astronomy,*

[3]*Department of Chemistry Iowa State University, Ames, IA 50011, U.S.A*

[4]*Department of Physics and Center for Nanointegration CENIDE, University of Duisburg-Essen,*

*Lotharstrasse 1, 47057 Duisburg, Germany*

## ABSTRACT

The realization of the unusual properties of 2-D materials requires the formation of large domains of single layer thickness, extending over the mesoscale. It is found that the formation of ideal graphene on SiC, contrary to textbook diffraction, is signalled by a strong bell-shaped component (BSC) around the (00) and G(10) spots (but not around the substrate spots). The BCS is also seen on graphene grown on metals, because a single uniform graphene layer can be also grown with large lateral size. It is is only seen by electron diffraction but not with X-ray or He-scattering. Most likely it originates from the spatial confinement of the graphene electrons, within a single layer. This leads to large spread in their wavevector which is transferred by electron-electron interactions to the elastically scattered electrons to generate the BSC.

**Corresponding author : mctringi@iastate.edu

Graphene has been intensively studied as a novel 2-d material because of its unique band structure, with potential graphene applications predicted in many technologically important areas[1-4]. The key goal is to grow graphene of the highest quality, i.e., of uniform thickness, and lowest density of defects. Similar goals have become a current priority for the growth of other 2-d van der Waals bonded materials with electronic band structures similar to graphene[5]. In this letter we demonstrate that a surprising result can be used to identify the optimal conditions (temperature, annealing time) for graphene growth: paradoxically a very broad bell-shaped component (BSC) emerges around both the specular (00) and the graphene G(10) spots, signaling the formation of the perfect layer. Although this component has been seen in numerous other experiments in the previous literature, it has been ignored and has not been correlated with graphene uniform growth [6-11]. The component's FWHM is as large as 50% of the surface Brillouin zone (BZ) and since in diffraction broad peaks correspond to disorder, it is intriguing that it signals a highly uniform film. This conclusion is based on the unusual dependence of the BSC on electron energy found in the current experiments. They show that the BSC is not related to the scattering condition changing from constructive to destructive interference between adjacent terraces[12]. Instead it is a consequence of the single layer graphene that confines the electrons with very high precision normal to the surface. As expected and as seen in ARPES experiments the spatial confinement causes a large spread in the normal component of the electron wavevector $\Delta k_z$[13]. The confinement extends coherently over mesoscale distances, since graphene overgrows substrate steps. The effect is unusually strong, fundamental and general that it should be present in other 2-d van der Vaals materials, of similar single-layer uniformity over the mesoscale[14].

The experiments were performed on 4H-SiC(0001) purchased from Cree, Inc. The samples were graphitized in UHV (P~1×10$^{-10}$torr) by direct current heating of the sample to ~1200°-1400° C. Spot Profile Analysis Low Energy Electron Diffraction (SPA-LEED) is used for the measurements, with its higher reciprocal space resolution allowing quantitative analysis of the patterns [12]. Since elastically scattered electrons are collected within 0.5eV below the beam energy, the BSC does not originate from plasmons [15] that involve higher energies[7]. The transition from the buffer to monolayer graphene is described in terms of the evolution of a small number of spots: the 6x6 spots around all fundamental spots and a 3-spot cluster (close to the 1/3,1/3 position along [1$\bar{1}$00]). Fig.1 shows a 2-d diffraction pattern of the surface partially covered with buffer layer(BL) and single layer graphene (SLG). The spots in the cluster (the 5/13, along [1$\bar{1}$00] and the two neighboring spots along [1$\bar{2}$10]) are attenuated as graphene grows with further annealing. Since the BSC is not seen around the SiC spots, this implies that it originates from graphene electrons. Although the BSC and the spot evolution towards single layer graphene (SLG) were seen before in the listed patterns[6-11], they were not mentioned. In the current experiments spot profile measurements, as a function of energy were carried out to understand the origin of BSC and its relation to graphene morphology.

Graphene growth on the Si-face of SiC is carried out at high temperatures (above ~1200° C) so Si evaporates while the remaining C diffuses and forms a uniform layer. Within a 200° C window the grown thickness changes progressively from buffer layer (BL) seen by the growth of (6√3x6√3), to single-layer, to bilayer and few-layer graphene. The earlier work shows that the BSC evolves as the substate changes from initial $6\sqrt{3}x6\sqrt{3}$ to thick graphene. It starts appearing around the (00) spot after annealing at 1200° C. With temperature increase the spot 5/13 characteristic of the BL disappears (indicating the formation of SLG) while the BSC becomes stronger. Although the previous experiments have captured slightly different snapshots of the graphitization process, they are in agreement that BSC is a measure of graphene uniformity. Onset of BSC around G(10) with $\sqrt{3}x\sqrt{3}$ and $6\sqrt{3}x6\sqrt{3}$ phases coexisting is seen in [6,10]. More intense BSC is seen when SLG forms with the BSC starting to decrease after bilayer [7] or graphite form [2] when the sample is heated to higher temperatures. The full evolution from BL to few layer graphene studied in [8] shows that the BSC is maximal in the middle of the temperature range when SLG is present. The correlation between strong BSC and high quality graphene was also seen in the current experiments. Fig. S1 shows the onset of graphitization (black curve at 1200° C ) and after the completion of a SLG (green curve at 1300° C). The BSC (shaded areas around (00) and G(10)) is stronger when SLG is completed. The electron energy is 194eV. Quantitative analysis shows that the normalized BSC grows 3 times around (00) spot, 6 times around the G(10) and the normalized area of the 5/13 decreases by a factor of 5.

Fig. 2(a) shows 1-d scans at E=148eV of the (00) spot (fig.2(a) along [1$\bar{2}$10]) and fig.2(b) at 132eV along [1$\bar{1}$00]) with G(10) seen. The high resolution of SPA LEED is a clear advantage (over normal LEED) because it shows two distinct components of the (00) spot (while in refs. [6-11] this is not possible). The narrow component has FWHM= 0.5%BZ, and the BSC has FWHM=33%BZ. In textbook diffraction, broad spots commonly imply the presence of disorder and non-uniformity on the surface. However the profiles of fig.2 are very unusual because the broad components have FWHMs, which correspond to a distance as small as $\sim 2a_g$, with $a_g=0.245nm$ the graphene lattice constant. All studies of graphene with different probes have not identified any feature at this short length scale. Fig.2(b) shows the BSC around the two G(10) spots, the FWHM of the narrow component of G(10) is 2.25%BZ and the FWHM of the BSC of G(10) is smaller than the FWHM of the (00) spot at the center (by 20%).

Fig.3(a) shows in a pictorial way the spot profiles as a function of $k_{\parallel}$ over a range of energies 100-200eV. The intensity maxima are at 104eV, 144eV, and 200eV, surprisingly at the same energies both for the narrow and BSC components; correspondingly the minima are at 124eV, 160 eV again at the same energy for both components. This paradoxical result by itself suggests that the origin of the BSC is not related to changes of the scattering condition between adjacent terraces, from destructive to constructive interference [12]. If this was the case the narrow component should be anti-correlated to the BSC component, i.e., when the narrow component reaches a maximum (i.e. constructive interference) the broad component should reach a minimum (i.e. destructive interference). Fig.3(b) shows profiles of the G(10) spot as a function of energy over the same range. Maxima and minima are correlated to each other as for the (00) spot; although shifted to lower energy from the extrema of fig.3(a) by approximately $\sim 15eV$.

Fig.4 shows the integrated areas of the (00) narrow component $A_{nar}$ (cyan), of the BSC background $A_{bro}$ (blue) and their normalized ratios $R_{00}= A_{nar}/( A_{nar}+ A_{bro})$, confirming again the correlation. The energy is shown in the top and the reduced variable $s=\Delta k_z/2\pi/d_g$ at the bottom scale (where $\Delta k_z$ is the momentum transfer normal to the surface). In addition the three maxima are close to half integer values of s=5.5, 6.5, 7.5, while if the BCS was due to scattering from adjacent terraces, the maxima of $A_{nar}$ should be for integer values n so the phase shift s is $2n\pi$ with n an integer [12].

Epi-graphene (EG) can be grown on either of the two polar faces of SiC, the SiC$(000\bar{1})$ (C-face) or SiC (0001) (Si-face). Graphene grown on the Si-face of SiC is more uniform and extends to large lateral size. It has been extensively used to study its electronic and topological properties of graphene[2,3,16] and more recently to grow 2-d materials by intercalation[17,18] .

On the other hand graphene grown on C-face has a larger number of layers (more than ~10) and the domain sizes are smaller [16]. The BSC is only seen on Si-face graphene because of the larger domain size and single layer thickness. No BSC is seen on graphene grown on C-phase of SiC, which confirms that high density of defects can destroy the BSC. But a smaller density of defects (seen in preliminary low coverage metal deposition experiments) shows essentially intact BSC. The quality of graphene when the BSC is present is seen in all studies [7,8,11]. Earlier STM experiments show that graphene domains reach micron sizes [19]. Recent characterization with three complemenatry techniques (SPA-LEED, STM and ARPES), confirm the graphene high quality from the presence of strong replica Dirac cones [20].

The spread $\Delta k_z$ of the graphene electrons confined in graphene of uniform thickness can be transferred to the diffracted electrons during scattering via electron–electron interactions. Because the scattering is elastic this can generate a spread to the parallel component $\Delta k_{\parallel}$ of the scattered electrons which can be expressed

$$\Delta k_{\parallel}=-k_z\Delta k_z/k_{\parallel}=-(E-(h^2/2m_e)(k_{\parallel}^2))^{1/2}(1/d_g)/k_{\parallel} \qquad \text{eq.(1)}$$

where $(\Delta k_z, k_{\parallel})$ define the momentum transfer spread for the spot under investigation and $m_e$ the electron mass. This transfer is a likely mechanism generating the BSC.

This type of scattering is unique to graphene (and not to other ultrathin films) because of the large continuous domains, which overgrow even substrate steps, like a carpet. Graphene is the only system showing BSC. The electron confinement condition is better satisfied for laterally larger "infinite" domains. Graphene (and other 2-d materials) fulfill such conditions because they have single thickness and are not interrupted by steps. The BSC is a good measure of thickness uniformity and lateral size. It becomes visible when the initial small BL domains become comparable to the coherence length of video LEED ~30nm[6-11], so it is a great diagnostic over a wide range of graphene domain sizes. In all other cases films are interrupted at a step, which limits the spatial extent of the electron wavefunction in the film and the coherency in scattering between the incoming and valence electrons. As multilayer graphene of height $md_g$ grows with further annealing, electron confinement is reduced and the FWHM $\Delta k_z \sim 1/md_g$ decreases with m, consistent with the stronger BSC when monolayer graphene is grown.

Another observation supporting the previous result relates to graphene grown on metal surfaces, since this type of graphene is also highly uniform and overgrows steps, so BSC is seen in such films. For graphene on Ir(111) graphene forms by the thermal decomposition of ethylene above 1400° C with ethylene pressure of $5 \times 10^{-6}$ mbar. Only the oriented R0 phase is present indicating highest quality of graphene. The BSC is similar with FWHM ~50%BZ, seen both around (00) and Ir(10) spots (fig. 1 of ref.[21]. A Moire pattern also forms with 10 spots between (00) and Gr(10) (because $9a_{Ir} \approx 10a_g$). Because on this surface the graphene growth is along the Ir(111) unit cell (while on SiC is rotated by 30° from the SiC unit cell), it is seen that the BSC around Ir(10) is centered not on Ir(10), but on the Gr(10) spot. This confirms that graphene electrons is the cause of BSC ( the G(10) rod is further away from (00) than the Ir(10) rod ).

The proposed mechanism can explain more experimental observations in the literature. As noted increasing $k_{\parallel}$ and fixed E (i.e., comparing the G(10) vs the (00) spot), the FWHM decreases with $k_{\parallel}$ as predicted from the smaller ratio $\Delta k_z/k_{\parallel}$ for the graphene spot (in eq. (1)). With increasing beam energy E and for a given spot (so $k_{\parallel}$ is fixed) the BSC overall increases with energy, as expected from eq.(1)[15].

More information can be obtained about the BSC by comparing scattering experiments using different probes. From the early graphene studies it was noted that X-ray scattering on graphene grown on SiC shows only a single narrow component both on Si-and C-phase graphene as seen in ref. [22]. Similarly He-scattering experiments on graphene grown on Pt(111) also show one component profiles with FWHM similar to the clean Ni(111) substrate[23]; this indicates that only long range order is probed in the X-ray and He-scattering experiments and no BSC is present. The interaction between the graphene electrons (which have large $\Delta k_z$), with either the photons in the X-ray beam or the He-atoms in the He beam, is much weaker so there is no transfer of this large momentum spread to the diffracted beam. Besides the LEED experiments on different types of graphene, the BSC has been also seen in experiments with $\mu$-LEEM[24] and with RHEED[18]. $\mu$- LEEM was performed on graphene grown on Pt(111), with domains consisting of a range of Moire supercells and of graphene domain sizes as large as ~50 $\mu$m (generating large $\Delta k_z$). The BSC was also seen in RHEED experiments studying superconductivity of intercalated graphene on SiC with Ca[18].

A large number of diverse observations has been presented that show the BSC to be a very strong feature in electron scattering from graphene. It is present in all previous growth experiments that show very high quality graphene. It clearly relates to the unusual graphene single layer uniformity that can extend without interruptions beyond mesoscale distances.

It is a very robust feature but its origin very intriguing and unconventional. A plausible physical mechanism generating the BSC that relates to the unprecedent graphene uniformity and can account for all observations relates to electron confinement. The position of the graphene electrons is very precisely known within a single layer $d_g$=0.33nm, so they have a large variation in their wavevector normal to the surface, as determined by the uncertainty principle $\Delta k_z > 1/d_g$ (i.e., $\Delta p_z = \hbar/d_g$). The incoming electron beam primarily interacts with the atomic core (the atomic scattering factor is determined by the charge distribution of the protons in the C nucleus and the surrounding electron clouds in the C atoms). Electron-electron scattering between the incoming electron wave and the graphene valence electrons can also play a role (in the change of the electron wavelength due to the inner crystal potential and scattering resonances in the image potential, as discussed in scattering textbooks). Because of the elastic character (E=constant) of the diffraction process (irrespective of whether the incoming beam interacts with graphene atoms or valence electrons) the undefined value of $\Delta k_z$ in graphene electrons is transferred to the elastically scattered electrons.

In conclusion an unusually broad background (BSC) seen in electron scattering experiments was studied quantitatively. Paradoxically it signals the formation of perfect graphene, contrary to textbook description that broad features in diffraction indicate disorder. Detailed studies of the diffraction profiles with energy rule out the standard explanation in terms of the variation of the scattering phase from constructive to destructive interference. The BSC is seen only around the (00) and G(10) spots but not around the SiC spots, it is seen only on the Si-face and not the C-face graphene because of the larger, uniform domains grown on the former; and it is seen for graphene grown on metals. BSC is not seen in X-ray or He-scattering scattering experiments. Its origin was attributed to the spatial localization of the graphene electrons, within a single layer, when ideal graphene is completed. This results in large spread in the wavevector normal to the surface $\Delta k_z > 1/d_g$ as a result of the uncertainty principle. This spread most likely is transferred to the elastically scattered electrons through electron-electron scattering. Besides being a great diagnostic tool and a mesoscale realization of fundamental quantum mechanical effects it should be also present in other 2-d systems of current interest, also signaling perfection in their growth.

Acknowledgments.

This work was supported by the U.S. Department of Energy (DOE), Office of Science, Basic Energy Sciences, Materials Science and Engineering Division. The research was performed at Ames Laboratory, which is operated for the U.S. DOE by Iowa State University under contract # DE-AC02-07CH11358. Some of the figures were included in the PhD Thesis of M. T. Hershberger  Iowa State University.

**Figure captions**

Fig. 1 Diffraction pattern for mixture of buffer layer(BL) and single layer graphene (SLG) at energy E=194eV. The BSC forms around the (00) and G(10) but not the SiC(10) spots. Several spots are marked including the 5/13 spot and the two neighboring spots forming a 3-spot cluster. Its evolution tracks the transition from BL to SLG.

Fig.2(a) 1-D scan of the specular spot along the SiC direction [1$\bar{2}$10]), at E=148 eV. The FWHM of the narrow component is 0.5%BZ, of the BSC is 33% BZ which corresponds to a distance ~3$a_g$.The ratio of the integrated narrow to sum of narrow and BSC areas is ~0.65. Fig. 2(b) 1-D scan of the specular along the graphene direction, and E=132 eV. The FWHM of the G(10) narrow component is 2.25%BZ and of the BSC component is 80% of the FWHM of the (00) spot at the center of the scan. The integrated areas ratio of the narrow to the sum of the areas of both components is 0.5.

Fig. 3(a) 1-D profiles of the 00 spot collected every 4eV from 100 eV to 200eV. The color range is shown to the right (from $2x10^6$ to $10^3$). Fig. 3(b) 1-d profiles of the G(10) spot collected every 4eV from 100 eV to 200eV. The color range is shown to the right (from $1x10^5$ to $5x10^2$). The maxima are shifted with respect the maxima of the (00) in fig.3(a) because of the contribution of the non-zero parallel wavevector component of the G(10) spot. For both spots the maxima of the narrow and BSC components follow each other which is not consistent with scattering interference from adjacent terraces as the origin of BSC. (The initial bending of the Gr(10) spot is related to the non-linearity of SPA-LEED at the lower energy).

Fig. 4 The integrated areas of the narrow component (deep blue) and the BSC (light blue) plotted as a function of the scaled momentum transfer s=$(\Delta k_z)/(2\pi/d_g)$ (shown at the bottom, with the corresponding energy at the top). The two areas have the same variation with energy while they should be anti-correlated if the origin of BSC was tewxtbook scattering. The fraction of the narrow component defined by the ratio $R_{00} = \frac{A_{nar}}{A_{nar}+A_{bro}}$ is plotted in black with the maxima close to half integer values of s=5.5, 6.5, 7.5. Maxima are expected for integer values of s if the BSC was originating from interference between adjacent terraces.

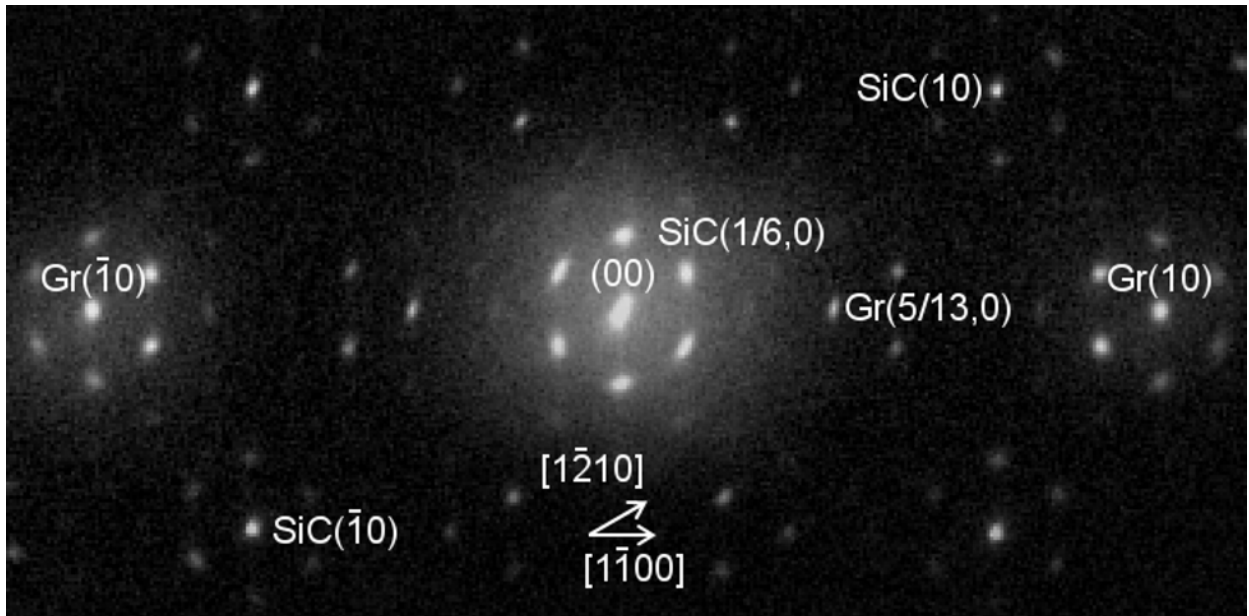

fig.1

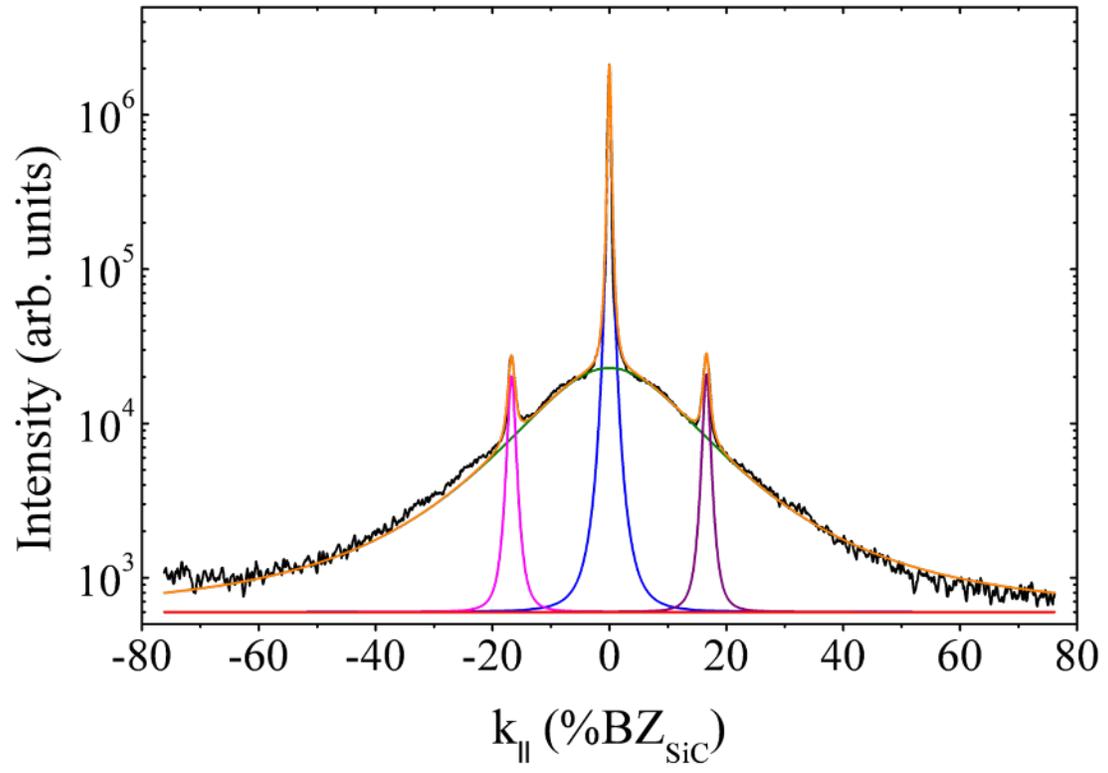

fig. 2(a)

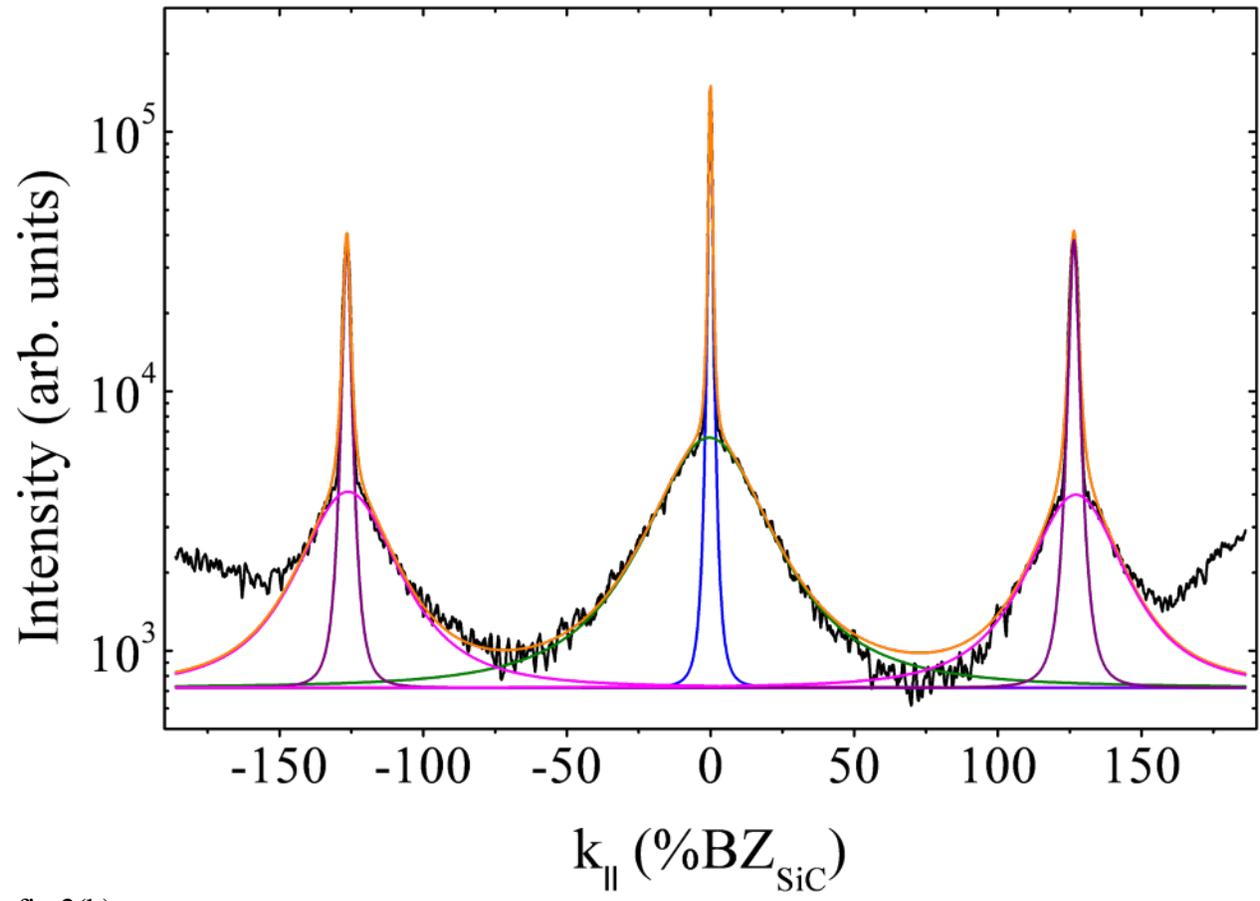

fig.2(b)

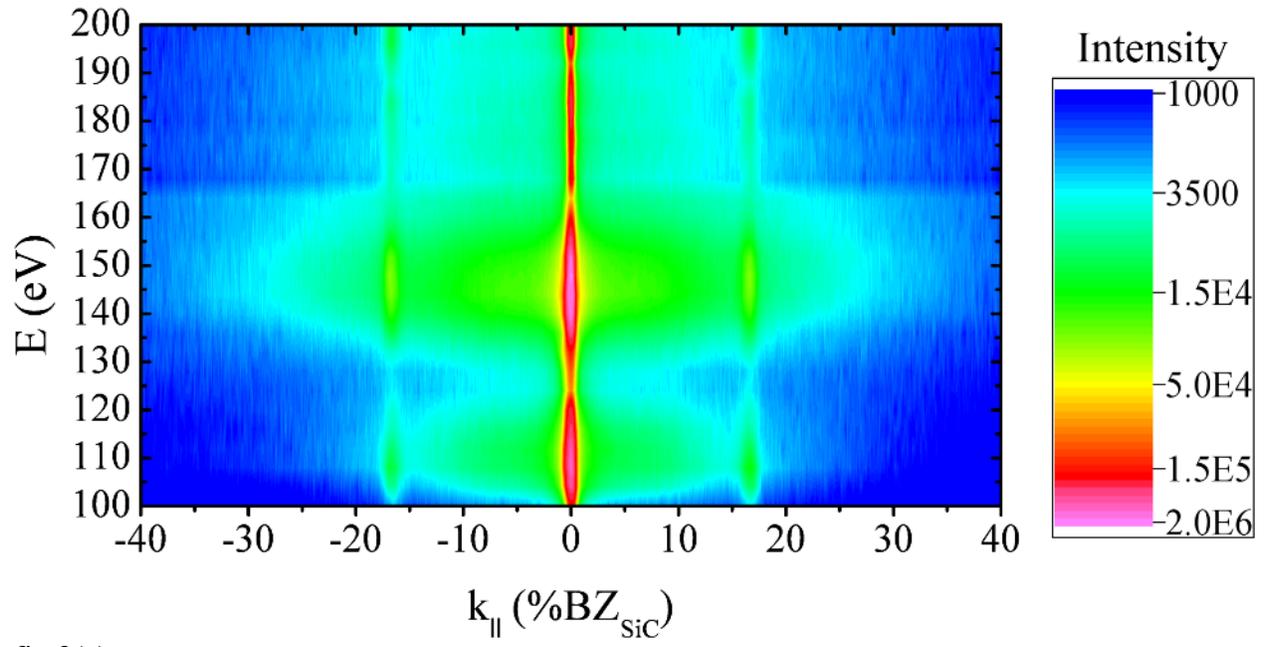

fig.3(a)

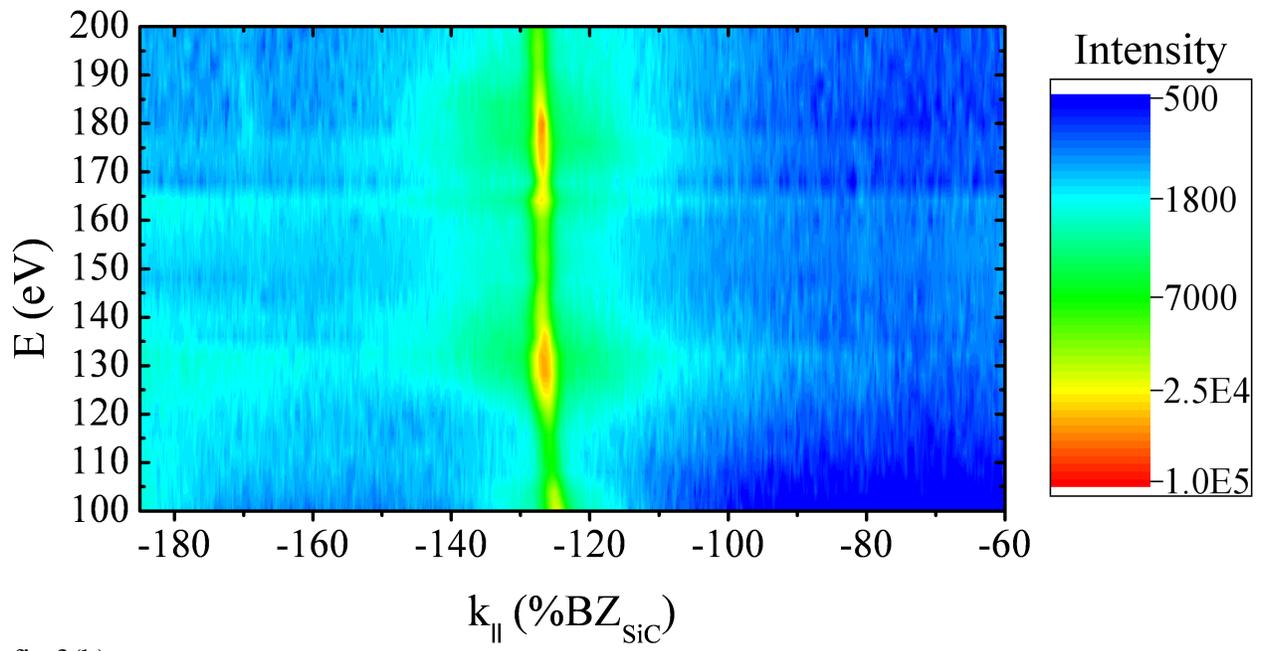

fig.3(b)

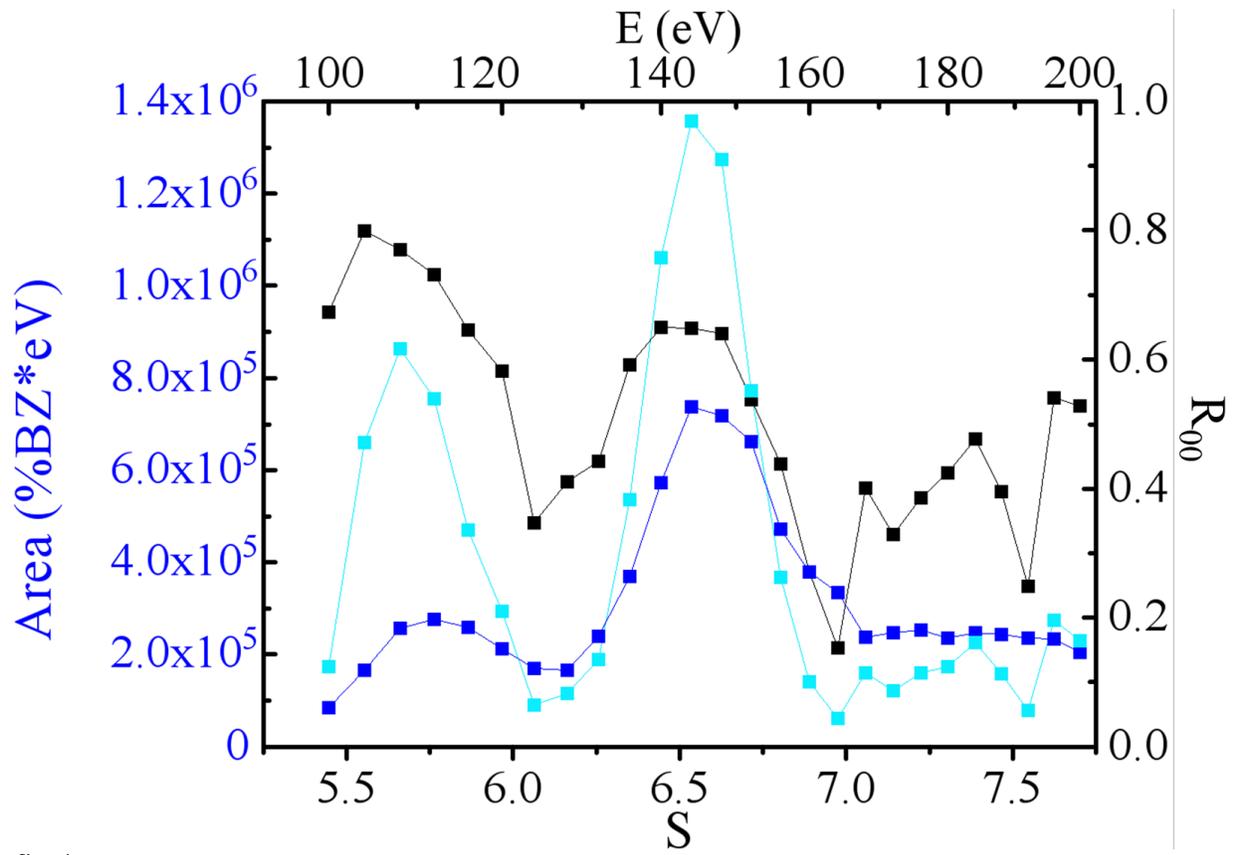

fig.4

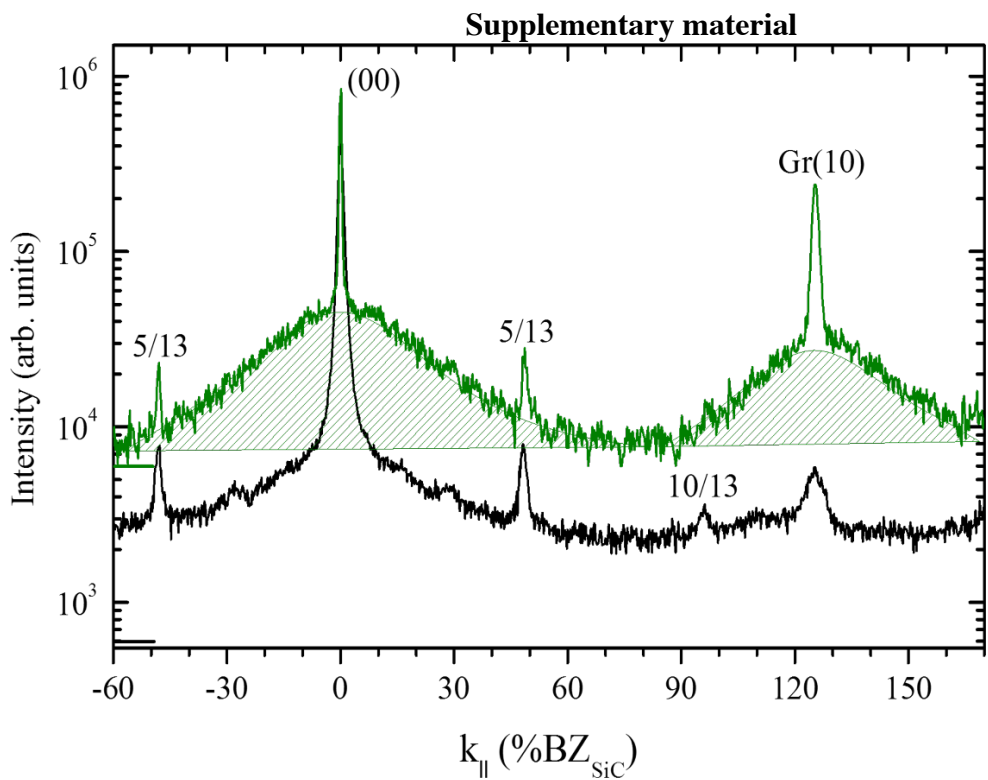

Fig. S1 Onset of graphitization (black curve at 1200° C ) and after the completion of a single graphene

layer (green curve at 1300° C). The BSC (shaded areas around (00) and G(10)) is stronger when single layer graphene is completed. The green curve is shifted for clarity and the markers to the left scale show the same intensity at 650 counts. The relative contribution of the BSC to the integrated area over the shown BZ increases by a factor of 3, of the 5/13 decreases by a factor of 5, of G(10) increases by a factor of 6 after annealing. This shows that the BSC correlates with the growth of SLG.